# Improving Resource Allocation in Software-Defined Networks using Clustering

Mahdi Sarbazi [1], Mehdi SadeghZadeh [2,*], Seyyed Javad Mir Abedini [3]

[1] Department of Information Technology Engineering, South Tehran Branch, Islamic Azad University, Tehran, Iran

[2,*] Department of Computer engineering, Mahshahr branch, Islamic Azad University, Mahshahr, Iran. Email:Sadegh_1999@yahoo.com. Phone:+989121406562

[3] Department of Computer engineering, Center Tehran branch, Islamic Azad University, Tehran, Iran.

**Abstract**

Software-defined networks (SDNs) are a huge evolution in simplifying implementation and network operation which have reduced costs and made the network programmable. Although SDNs are a suitable option for solving some of the previous problems, but they have some challenges as any other new technology. Resource allocation and balance control in network is one of the main challenges of this technology which is studied in this paper. In this study, a new approach is proposed for improving memory resource allocation in network using load distribution clusters. Since in the proposed method, K-mean++ algorithm is used for clustering, load balancing of clusters can be used to preserve load balance of the network. In the proposed method, data with higher recall is transmitted to high-quality clusters in terms of average number of hubs and lower average delay between server and user. In the proposed method, by increasing number of clusters, higher memory is created in the network.

***Keywords***: Software-Defined Networks, Resource Allocation, Clustering, Load Balancing.

# 1. Introduction

Using software-defined networks opens the way for new and innovative applications ahead of admins. Using this method, Permanent control of network topology becomes possible, access control in the whole network and energy management can be managed at any time. In addition, programmability of software-defined networks provides the possibility of continuous connection at all levels. This reduces resource consumption and admins can use network capacity better. Software-defined networks (SDNs) aim to simplify using network and reduce management costs using programmable services [1] [2].

Architecture of SDNs is comprised of application, controller and infrastructure layers from top to bottom as shown in figure 1. Infrastructure layer of the network is the bottom layer of architecture in which physical devices including routers and switches are located [2]. Controller layer is the middle layer of SDNs which is a software controller distributed on one or several servers. Connection of this layer with infrastructure layer is made possible through user interfaces. Operating Systems provide an objective layer for management of hardware resources and their security. This part is the most important part of SDN architecture which supports components of control logic. Among the most important controllers, POX, RYU, Floodlight and ODL can be mentioned.

Application layer is the top most layer of SDNs which includes a set of programs and softwares in network level. These programs and softwares announce their requirements through a series of middle interfaces to the controller. Conversation between controller and application layers is called northbound [3]. Conversation between controller and infrastructure is called southbound [4]. When controllers are distributed on different servers, conversation between these controllers is called east/west bound [5].

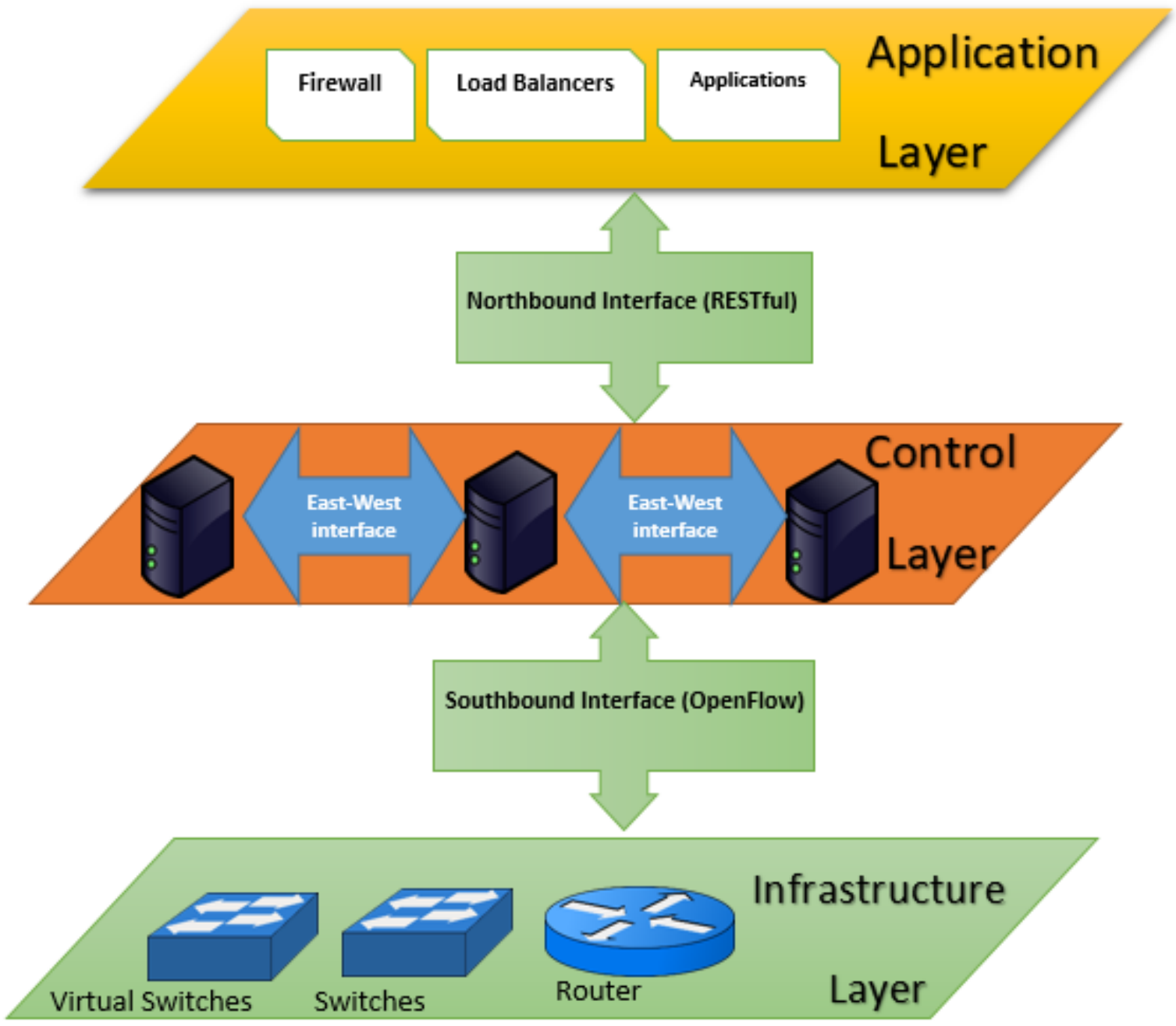

Figure1. Architecture of SDNs

Despite central controlling feature in architecture of SDNs, this architecture has some challenges while facing some problems. Resource allocation is a method which has the ability to optimize use of resources existing in the network. Although network operators use SDNs as an advantage by increasing control performance on the network but they cannot still respond requirements of large networks like flow request and network monitoring statistics. For instance, one of the old controllers called NOX can only accept 30k flow requests per second where its response time is less than 10 ms. This non-optimality in large networks can be seen compared to conventional networks [6].

Load balancing can be developed in software or hardware. Load balancing is responsible for load distribution among resources. A proper load balancing has various features like fast response time, maximum error tolerance and minimum usage of resources and preventing overflow [7]. There are two types of load balancing in SDNs: Load balancing in IP network and load balancing in SDNs. Load balancing in IP network is load balancing in infrastructure of SDNs which aims to balance network load among different distributed cluster servers existing in the IP network. This is done using commands transmitted from south bound of the networks and results in deciding about forwarding the packets such that load balancing is established.

Load Balancing in SDNs is load balancing in the controller layer of SDNs which aims to balance network load among distributed controllers or exchanges information in south bound and east/west bound of SDNs.

Purpose of this project is to present a method for improving resource allocation in SDNs based on clustering resources in infrastructure layer. This study is of load balancing in IP network type and employs resource clustering to improve resource allocation and increase storage capacity and load balancing.

The proposed method performs resource allocation aiming to increase data storage capacity and network loading balance. To this end, memory resources of the network are allocated using load distribution clusters. First, data stored in switch level and before loading information in the servers are pre-processed based on current status of the network to prevent a large number of servers with identical data and offer a proper solution for storing more data preserving loading balance. To this end, clusters of the servers are comprised and data is allocated based on center feature of the clusters (average feature of the clusters). In the proposed method, K-means clustering method is used. Preserving loading balance in clusters and assigning high-quality clusters in terms of number of hubs and lower average delay between server and user for data with more recall, quality of service and efficiency of the network are increased.

This paper is organized as follows. Section 2 reviews previous studies. Section 3 describes the proposed method. Section 4 evaluates the proposed method. Section 5 concludes the paper and finally, future suggestions are presented in section 6.

# 2. Literature Review

In this section, recent papers are reviewed in two section. The first section reviews load balance methods briefly and the second section reviews clustering methods.

## 2.1. Load balancing

As complexity of wireless networks has increased specially in 5th generation communication networks and wireless sensor networks, control and synchronization of them has become a great challenge. Wireless networks require isolation of data and control sections in future as what has occurred in SDNs. In order to manage high-scale wireless networks, load balancing challenge in several controllers should be solved.

Various methods have been proposed for load balancing. Best measure for classifying existing methods is their being deterministic or non-deterministic [7]. Non-deterministic methods include greedy, approximate and heuristic methods. Other methods are considered as deterministic methods.

Deterministic methods always generate the same output for a specific input. Their process is described using different equations [7]. Guo et.al. [8] have proposed a load variance-based synchronization (LVS) methods to balance load in multiple distributed controllers. Compared to previous synchronization methods which are the basis of this method, methods based on LVS operate like a real synchronization method until load of the server passes a specific threshold. This threshold is determined by minimizing overhead of each controller. Empirical methods have shown that LVS achieves a better efficiency in load balancing and results in better forwarding and lower overhead in synchronization. Load balancing method among multiple controllers in software-defined wireless networks which recall hybrid flows has been proposed by Yao et.al. [9]. In this algorithm, balance is achieved by distributed and centralized techniques. Network is divided to a set of clusters including several switches and controllers. Kang and Choo [10] have proposed an improved method in super SDNs which transfers flows of the network to a cloud environment. This approach is comprised of two sections including monitoring and decision making. This method has performed better than conventional methods in distributing work load in an environment by avoiding hunger. However, this method does not measure distributed load among servers and error tolerance. In [11], a method has been proposed based on response times of the server using flexibility of SDNs. This method has minimum response time and good load balancing. Although developing the mentioned method is simple. Its load balancing approach is optimal, low cost and highly scalable but it has not considered energy conservation. Furthermore, this method has used a controller which limits accessibility and scalability and creates bottlenecks.

Non-deterministic approach includes some algorithms which might behave differently in different executions of a specific input. These methods are used to find approximate-temporal solution which is a hard deterministic and costly solution [7]. Chou et.al. [12], have proposed a load balancing system based on Openflow which employs genetic algorithm. This system transmits data from user to a set of various servers based on load balancing strategy along with a predefined data flows. When a heavy traffic occurs or server’s load increases, the proposed method can help balance workload of a set of servers. Results have proved that this method has higher efficiency compared to conventional methods. This method employs a controller; thus, accessibility and scalability are not obtained and system has bottleneck error. Tu et.al. [13] have introduced a programmable middleware which can distribute traffic equally. This method employs design and application of SDNs to improve bandwidth usage and ensure quality of service. Purpose of middleware is to find optimal path for traffic in datacenter; thus, it employs a matrix called cost matrix to determine data transmission cost from one server to another. Since middleware does not operate based on specific features of a datacenter, it can be applied to existing datacenters. However, it has not considered quality of service constraints. Boero et.al. [14] have proposed an alternative method which can be implemented with small modifications to Beacon controller. The proposed method employs real time statistics of Openflow and Beacon for routing flows in some queues again to observe deadline time requirements or efficient queue balance in Openflow. According to multi-way greedy heuristic algorithm, all flows are first located in queue q0. Controller sorts a system based on estimated speed in a descending order. Order of the estimated speed is analyzed and each flow is allocated to queue based on minimum usability. Multi-way approach behaves satisfactorily and shows better efficiency and this standard has not changed. Wang et.al. [15] have designed a switch migration scheme for load balancing in SDN controller. First, load balancing information is collected by the monitoring module and then it is decided about switch migration. Then, an efficient load balancing model is constricted to preserve balance between migration cost and load balancing speed. Finally, an efficiency-aware scheme is created based on greedy methods. The proposed method with migration makes the SDN-based controllers reactionary and improves migration performance. However, they are not tested in large-scale real wireless environment and real traffic. In addition, delay and error tolerance of this method have not been investigated.

## 2.2. Clustering

The main goal of clustering is to partition a dataset into clusters in terms of its intrinsic structure without resorting to any a priori knowledge such as the number of clusters, distribution of the data elements, etc. Clustering is a powerful tool and has been studied and applied in many research areas including image segmentation [16] [17], machine learning, data mining [18] and bioinformatics [19] [20]. Various clustering methods have been proposed in recent decades, common clustering algorithms can be divided into four major types according to different accumulation rules [21]: hierarchical clustering, fuzzy clustering, partition-based method and density-based method. There is no general clustering algorithm that can handle any form of data. It is known that K-means clustering [22] is one of the simplest clustering methods used in recommendation systems because of its relative scalability and high efficiency. Liu [23] proposed an improved clustering recommendation algorithm using K-means method; this method clusters both the items and users and introduces a time decay function for pre-processing user ratings to increase recommendation accuracy.

It is well-known that K-Means clustering is highly sensitive to proper initialization. The classical remedy is to use a seeding procedure proposed by Arthur and Vassilvitskii which is known as k-means++ together with Lloyd's algorithm. In the seeding step of k-means++, the cluster centers are sampled iteratively using D2-sampling: First, a cluster center is chosen uniformly at random from the data points. Then, in each iteration of k iterations, a data point is selected as a new cluster center with a probability proportional to its distance to the already sampled cluster centers [24].

Spectral clustering, which exploits pairwise similarities of data instances, has been widely used in several areas such as image segmentation and community detection, because of its effectiveness to find clusters [25].

Sancheza et al. [26] introduced a novel neuro-fuzzy system FasArt (Fuzzy Adaptive System ART-based) for both clustering and classification purposes which was tested on a dataset produced by UNIPEN project. Motivation of the researchers behind this research work is to reduce the noise in image, face recognition, image compression, facial expression recognition, character recognition, steg analysis in image etc. [27].

Authors of [28] have presented a graph-theoretical clustering method based on two rounds of minimum spanning trees. Lin et al. [29] have presented a K-medoids algorithm based collaborative filtering (CF) that clusters the content features of resources to alleviate data sparsity using compressed user-behavior data. Meanwhile, many studies have proposed novel clustering algorithms based on CF to better deal with sparsity and scalability problem. Koohi et al. [30] have introduced a fuzzy C-means clustering method to improve performance of the user-based CF in sparse datasets. Birtolo et al. [31] have designed a clustering CF framework and two clustering CF algorithms, item-based FCCF (IFCCF) and trust-aware clustering based CF (TRACCF), to improve both the quality and the coverage of suggestions.

Authors of [32] have proposed a new scheme that makes full use of all rating information based on Kullback–Leibler (KL) divergence from the perspective of item rating probability distribution, and distinguishes different items efficiently when selecting the cluster centers.

Among fuzzy clustering methods, fuzzy c-means (FCM) is the most recognized algorithm. In FCM, an algorithm was presented which assumed that all features are of equal importance. In real applications, however, importance of the features are different and there exist some features that are more important than the others. These important features should basically have more effects than the other features in the formation of optimal clusters. So, in [33] an automatic local feature weighting scheme has been proposed which properly weights the features of each cluster. And, a cluster weighting process is performed to mitigate the initialization sensitivity of the FCM. Feature weighting and cluster weighting are performed simultaneously and automatically during the clustering process resulting in high quality clusters regardless of the initial centers.

In 1993, Banfield and Raftery first proposed model-based Gaussian (MB-Gauss) clustering, using eigenvalue decomposition of Gaussian covariance matrix to detect different cluster shapes. In [34], MB-Gauss has been extended to a fuzzy model-based Gaussian (F-MB-Gauss) clustering. However, performance of both MB-Gauss and F-MB-Gauss algorithms depend heavily on initializations and require a number of clusters to be assigned a priori. To solve these problems, an unsupervised learning schema for F-MB-Gauss clustering is proposed.

In the following, the proposed method for improving resource allocation and load balancing in SDNs based on resource clustering which is a type of load balancing in IP networks is described.

## 3. The Proposed Method

The proposed resource allocation method aims to increase data storage capacity and network load balancing. To this end, a preprocessing is performed for data clustering before load balancing. The proposed method preprocesses data stored in switches before uploading information in servers based on current position of the network to prevent large number of servers with the same data and present a proper solution for storing more data preserving load balance. To this end, clusters of servers are comprised and data is allocated based on characteristic of cluster centers (mean characteristic of clusters). Figure 2 shows flowchart of the proposed method.

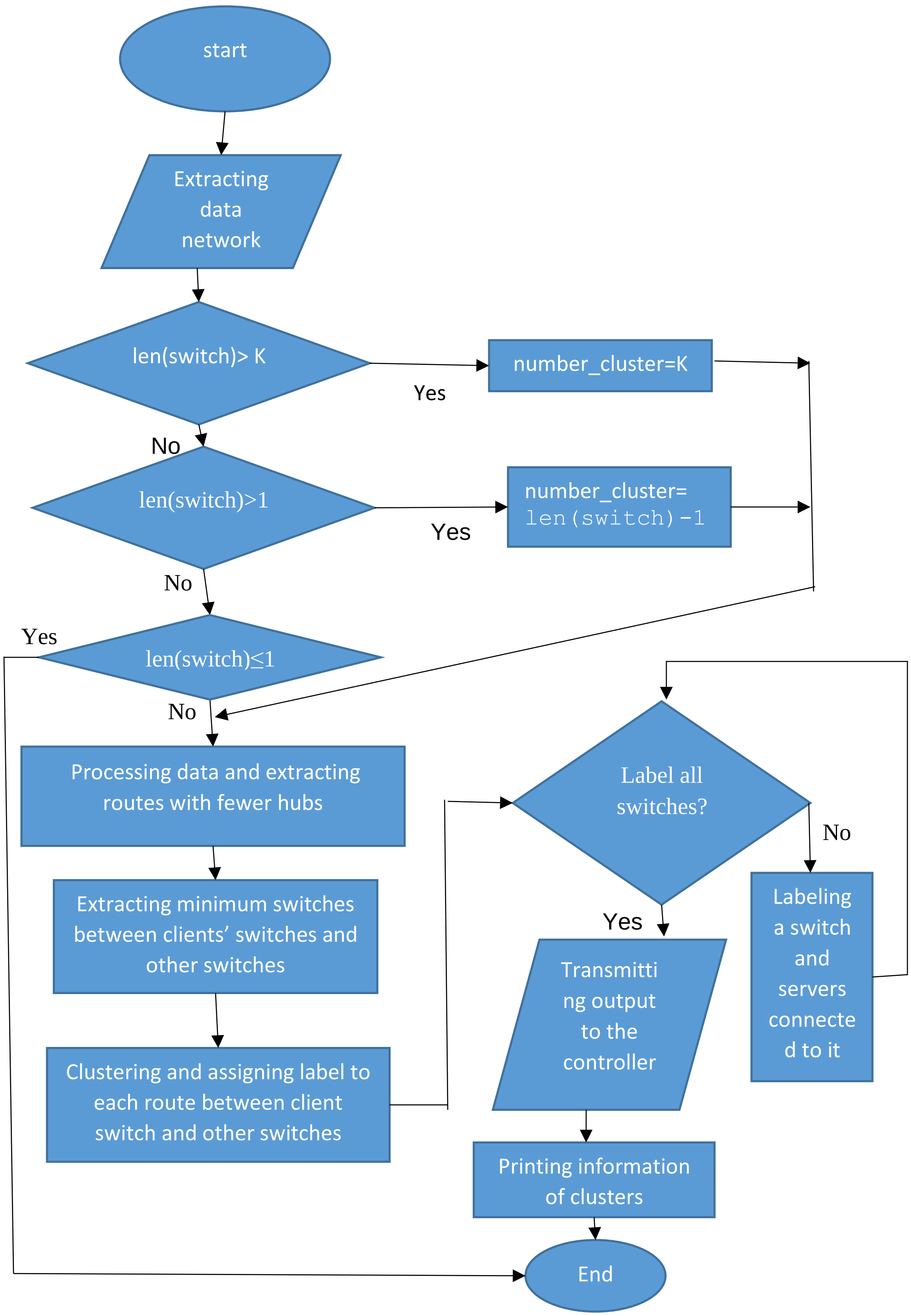

Figure 2. Flowchart of the Proposed Method

Input information is received from floodlight controller. If number of switches is more than k clusters, they are divided to k clusters. Otherwise, a number of clusters the same as number of switches (except switches specific to users' hosts) is defined. In the next step, routes with smaller number of hubs is obtained based on data extracted from topology. Then, minimum routes between user switch and server switches is selected among them. K-means++ algorithm is used for clustering. K-means++ algorithm in the proposed method calculates similarity measure based on number of hubs and delay. Then, servers connected to that switch are added to load balancing module based on labels assigned to each switch in clustering; finally, a number of clusters with specified characteristics are available using which network admin can connect a larger volume of data to each load balancing cluster based on characteristics of cluster head and data priority. Input and output data are recalled through REST API.

APIs are a set of subroutines and communication protocols and tools for building a software. In SDNs, different APIs in south bound, north bound and east/west bound can be used to process the results and if required, results are transmitted using methods defined for different parts of these networks to improve network performance. REST architecture style defines a set of constraints for using web service which creates cooperation possibility among different sections of computer system. This architecture is classified as web-based system APIs. This architecture style can be used in north bound of SDNs for writing applications. In this architecture, a URI which is a unique global ID is assigned to each resource or service or set of resources or services. This architecture is implemented using characteristics of HTTP protocol.

K-means++ algorithm is used for clustering. In this algorithm, value of k is considered as input and a set of n objects is partitioned to k clusters. Such that internal similarity of objects in clusters is maximum and similarity of objects outside clusters is minimum. Similarity in each cluster is measured using mean objects of the cluster which is also called cluster center.

In the proposed module, load balancing module in floodlight [35] is used to transmit output information. This module is a simple round robin load balancing module for icmp, tcp and udp protocols. This module is accessed through REST API. It should be mentioned that each load balancing method implemented in other controllers can replace this load balancer. But considering simplicity of using and advantages of floodlight controller, this controller is selected.

It is suggested to put datasets with higher recall probability in first clusters. Because initial clusters have fewer hubs and their delay is lower. Indeed, this scheme might be different based on requirement of network admin.

In similar server set, it is obvious that by increasing number of load balancing clusters, higher data capacity is obtained. Therefore, by increasing number of k clusters linearly, data storage capacity in the whole network increases k times. But, number of servers existing in each cluster is inversely proportional to number of clusters. Thus, by increasing number of clusters and considering that number of servers has decreased, quality of load balancing is reduced in normal state without clustering.

Among advantages of the proposed method k-times increase in stored data based on k clusters can be mentioned. Indeed, it should be noted that value of k should be selected properly with respect to n servers so that advantages of load balancing is preserved as much as possible. Purpose of this study is to employ clustering to improve balance between load balancing in data storage network and volume of stored data in the whole network.

# 4. Implementation and Evaluation

The proposed method includes a preprocessing on data clustering before load balancing to increase storage capacity in data networks. The proposed method is called K-means considering that K-means clustering is used. Evaluations are performed in similar environment with similar emulator, controller and configuration. In order to generate TCP type load, network uses iperf [36] and many of the information are obtained considering available outputs of this tool [36] . In order to implement k-means, library presented in [37] in Python programming language is used and inputs (information of minimum route extracted between two user switches and server) are defined in 2D where their dimension is number of hubs and delay.

In this evaluation, a 4-level topology is used. Each switch at each level is linked to all switches in its subsequent level, but they have no connection with switches in their own level; in each level, except first level, one switch with one host (server) and one switch with two hosts (servers) is located. Employed topology is given in the following figure.

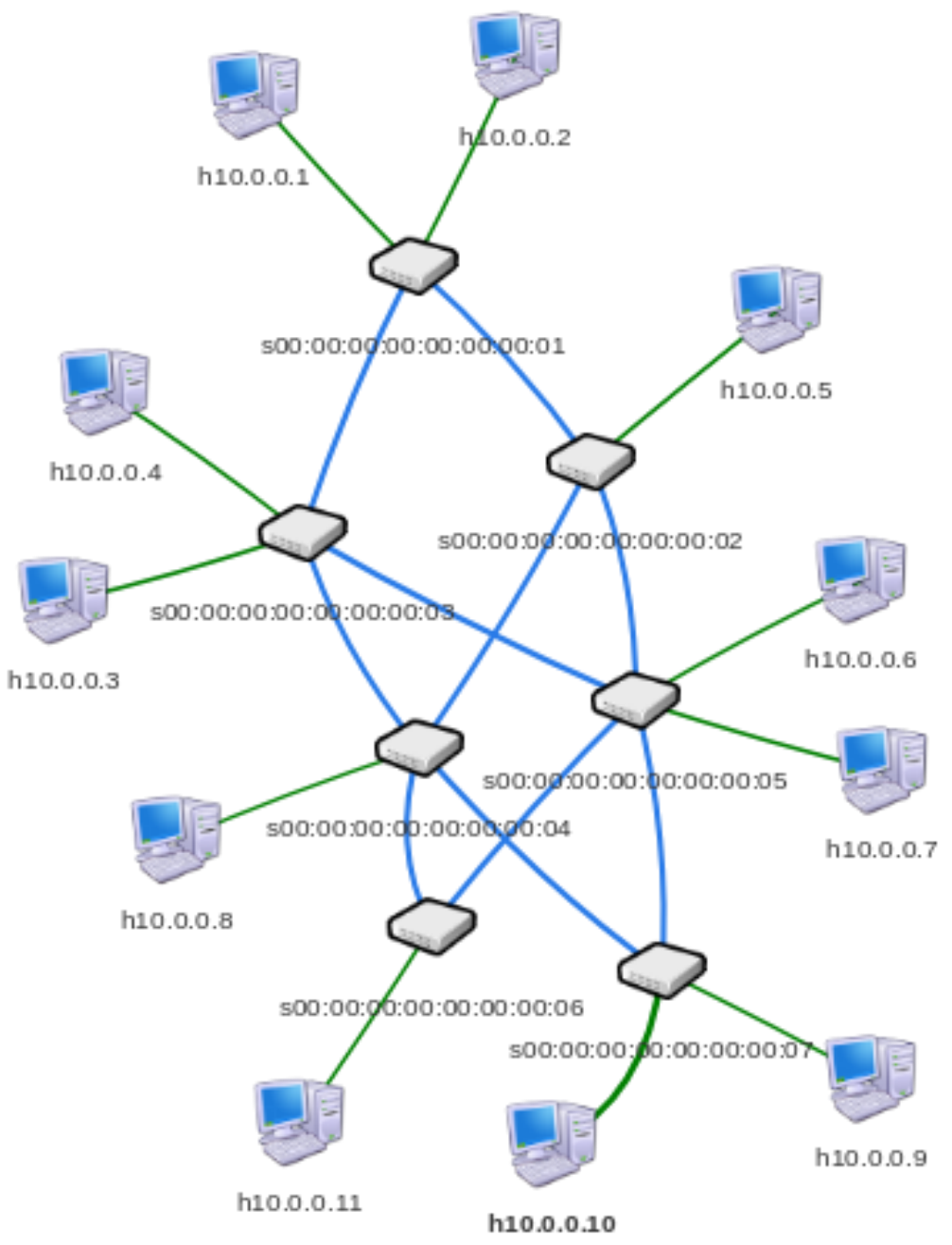

Figure 3. Network Topology

As can be seen in Figure 3, there is a letter along with IP of host or ID of switch. From now on, in order to read diagrams faster, name of switches or hosts is read by beginning letter or ending numbers after dot or colon; for instance, h10.0.0.1 is read h1.

In order to evaluate the proposed method, it is implemented using two clustering methods including k-means++ and spectral clustering

In the K-means++ method, servers are considered in three different clusters for clustering servers. As shown in figure 4, in the first cluster, average number of hubs between user and server is 1 and average delay calculated by controller in shortest route is 12ms. In the second cluster, average number of hubs between user and server is 3 and average delay calculated by the controller is in shortest route is 30.33ms and in the third cluster, average number of hubs between user and server is 2 and average delay calculated by the controller in the shortest route is 22ms. Figure 4 shows clustering result in the K-means++ method. While in Spectral Clustering method, clustering result would be as shown in figure 5.

In order to find optimal value of k in both k-means and spectral clustering algorithm, Table 1 is used:

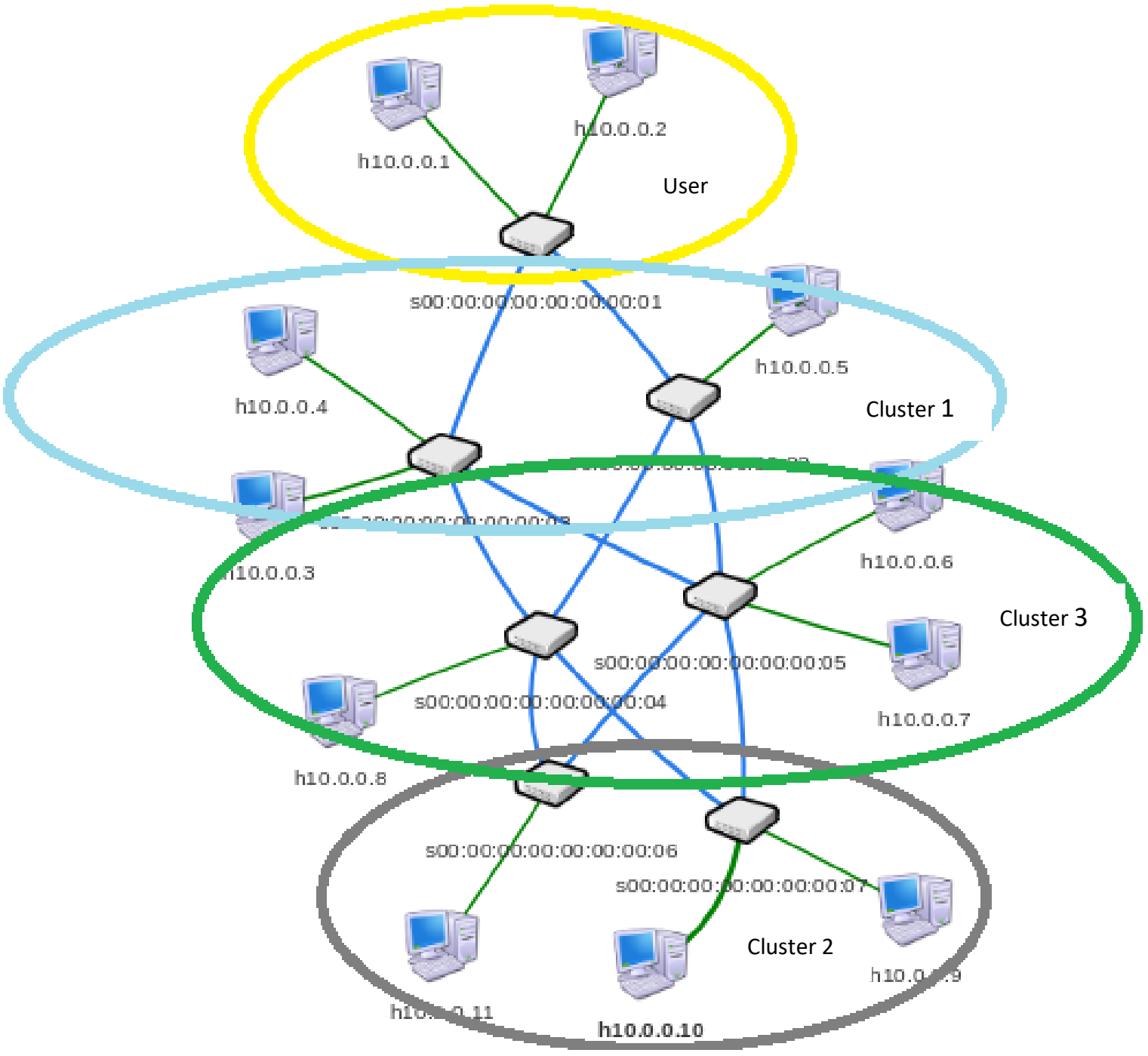

Figure 4. Clustering result in network topology in the K-means++ method

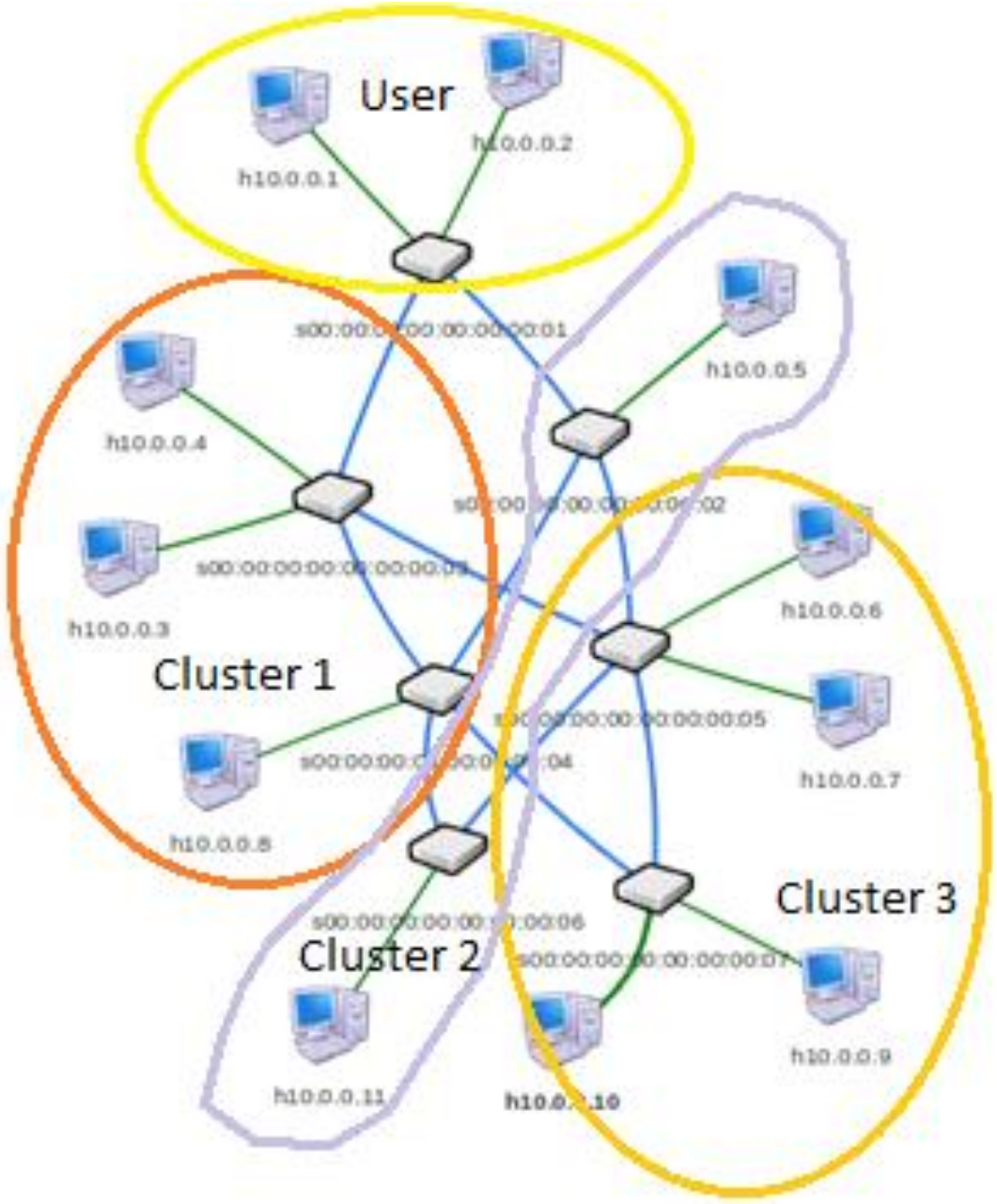

Figure 5. Clustering result in network topology in the spectral clustering method

Table 1. Comparing value of k in k-means and spectral clustering algorithms

| **Number of clusters (k)** | **1** | **2** | **3** | **4** | **5** | **6** | **7** | **8** | **9** |
|---|---|---|---|---|---|---|---|---|---|
| **Average of the servers existing in each cluster** | 9 | 4.5 | 3 | 2.25 | 1.8 | 1.5 | 1.28 | 1.125 | **1** |
| **The space added to data center before clustering** | 100% | 200% | 300% | 400% | 500% | 600% | 700% | 800% | **900%** |
| **Average load of 30 requests on each server in the largest possible cluster** | 3.33 | 1.875 | 1.428 | 1.25 | 1.2 | 1.25 | 1.428 | 1.875 | **3.33** |

In the third comparison, since selection of clusters is random, a load of $\frac{1}{number\ of\ clusters}$ is assigned to each cluster which is divided among all remnant servers of the largest cluster. Thus, its formula is:

$$A = \frac{\text{total head requests}}{\text{total number of clusters} * \text{total number of servers of the largest cluster}} \quad (1)$$

In this formula A is equal to Average load the number of requests on each server in the largest possible cluster. As can be seen from the table, cross cases are of high quality but when k=3, since it owns mean servers of all servers of the cluster, number of servers of each cluster should not be less than 3 so that load balance is preserved.

## 4.1. Time complexity

In order to analyze complexity of the proposed methods considering the flowchart presented in Figure 2, complexity of the proposed algorithm ignoring complexity of reading data from the controller and transmitting processed data for clustering for k-means++ and spectral clustering would be as follows:

For k-means++:

$$O(f(n)) = O(t\ n\ K) + O(n^3) \quad (2)$$

In this equation, n is the number of nodes existing in the network graph and t is the number of iterations of k-means algorithm and k is number of clusters $O(n^3)$. In order to find the shortest path between each two vertices of the graph, fluid-varshal algorithm is used. [38]

For spectral clustering:

$$O(f(n)) = O(n^3) + O(n) \quad (3)$$

In this algorithm n is the total number of the graph and $O(n^3)$ is complexity of the spectral clustering algorithm and O(n) is complexity of finding adjacency matrix as the input of spectral clustering algorithm.

## 4.2. Workload of each Server

In this evaluation, three different states are defined. In the first state, all servers are defined in a large cluster and 30 TCP requests are transmitted for set of servers. In the second state, three clusters are created by executing the algorithm and 10 TCP requests are transmitted for each cluster (assume one third of requests are given to each cluster). In the third state, there is a connection between two hosts with 30 simultaneous requests sent to a specific server. Following diagram shows number of requests in all three states from which workload of each server can be obtained considering number of requests.

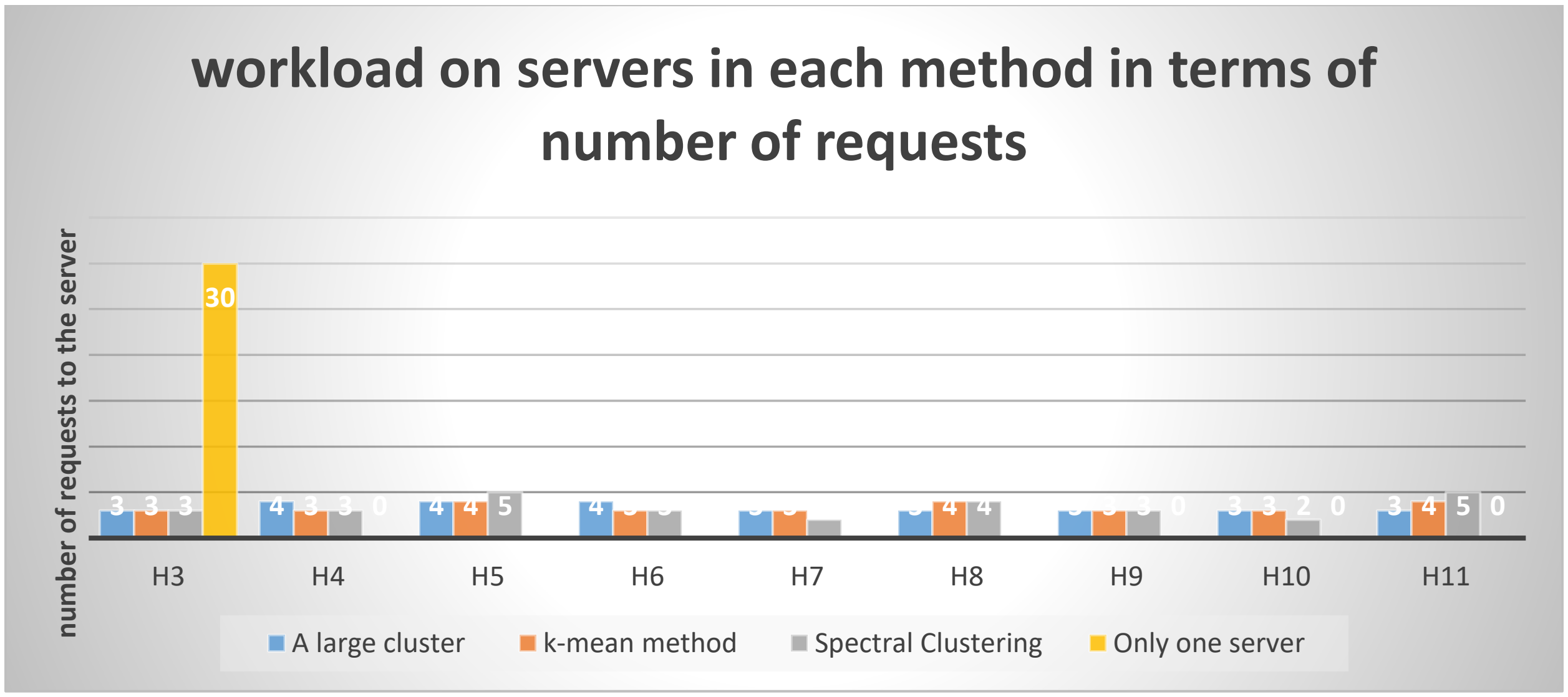

Figure 6. Workload of servers in each method considering number of requests.

As can be seen in Figure 6, in the first state, there is a high workload on h3 and not request has been sent to other clusters. In large cluster method, all requests are distributed equally among all servers. But, in the proposed method which is called k mean, requests distributed among servers are balanced and considering that number of servers in each cluster is the same, servers which are in the same cluster have experiences the same workload as their previous server. Thus, load is distributed similarly among all clusters.

### 4.3. Volume of Transmitted Data and Occupied Bandwidth in Servers on user side

In the conducted experiments, three states are considered. First 10 TCP requests are transmitted simultaneously to a large cluster including all servers and in the second state, the proposed method is employed to transmit 10 TCP requests to each cluster and in the third state, 30 simultaneous requests are transmitted to the large cluster as in state 1. Results of transmitted data volume and occupied bandwidth in server can be seen in Figures 7 and 8.

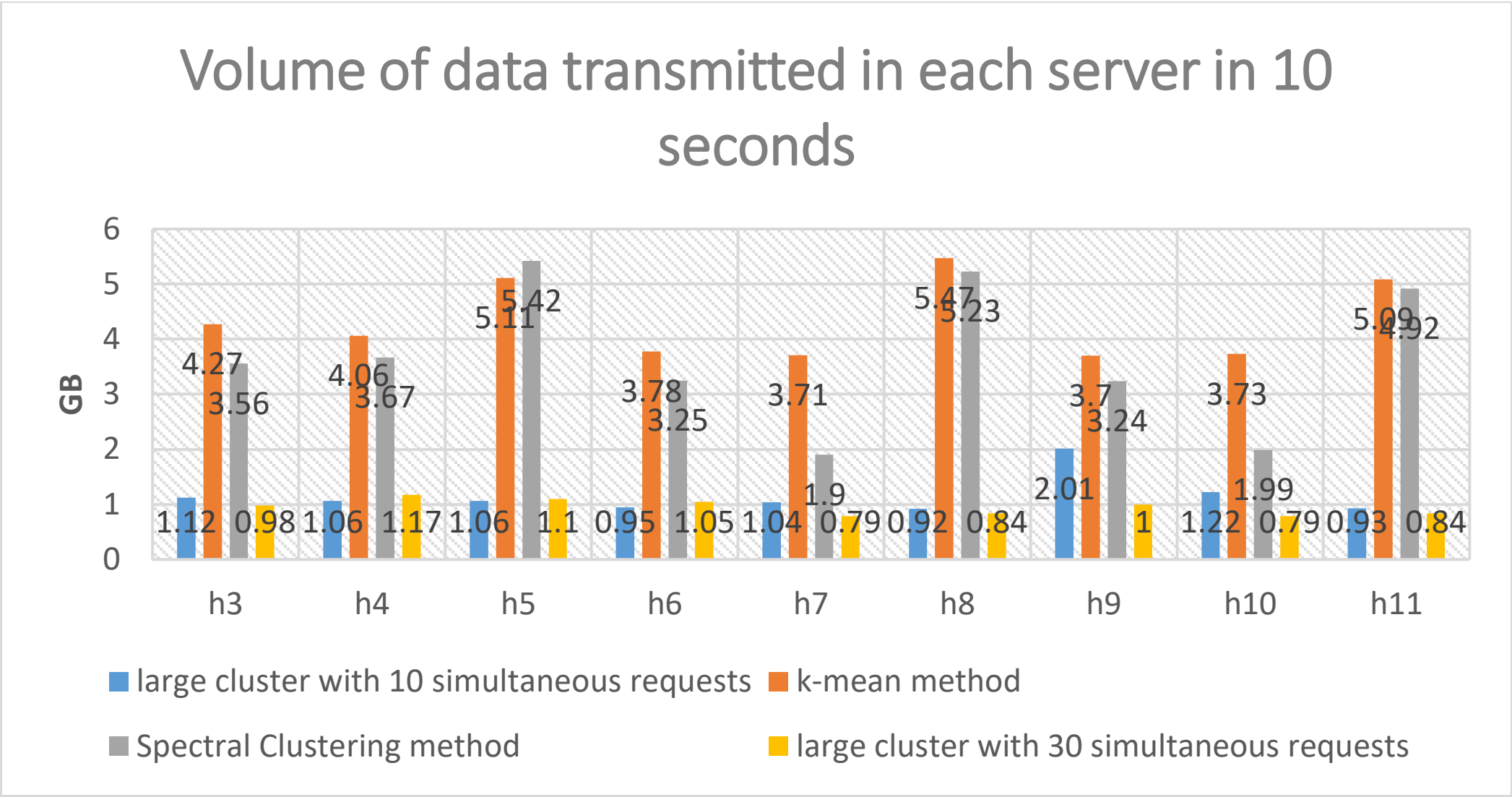

Figure 7. Volume of data transmitted in each server in 10 seconds

As shown in figure 7, Volume of Transmitted Data in k-mean and spectral methods is better than the other methods. Figure 8 shows that bandwidth usage on server's side in k-mean and spectral methods is higher than the other methods. Higher Volume of Transmitted Data and bandwidth usage on server's side show that our proposed method has performed better than other methods.

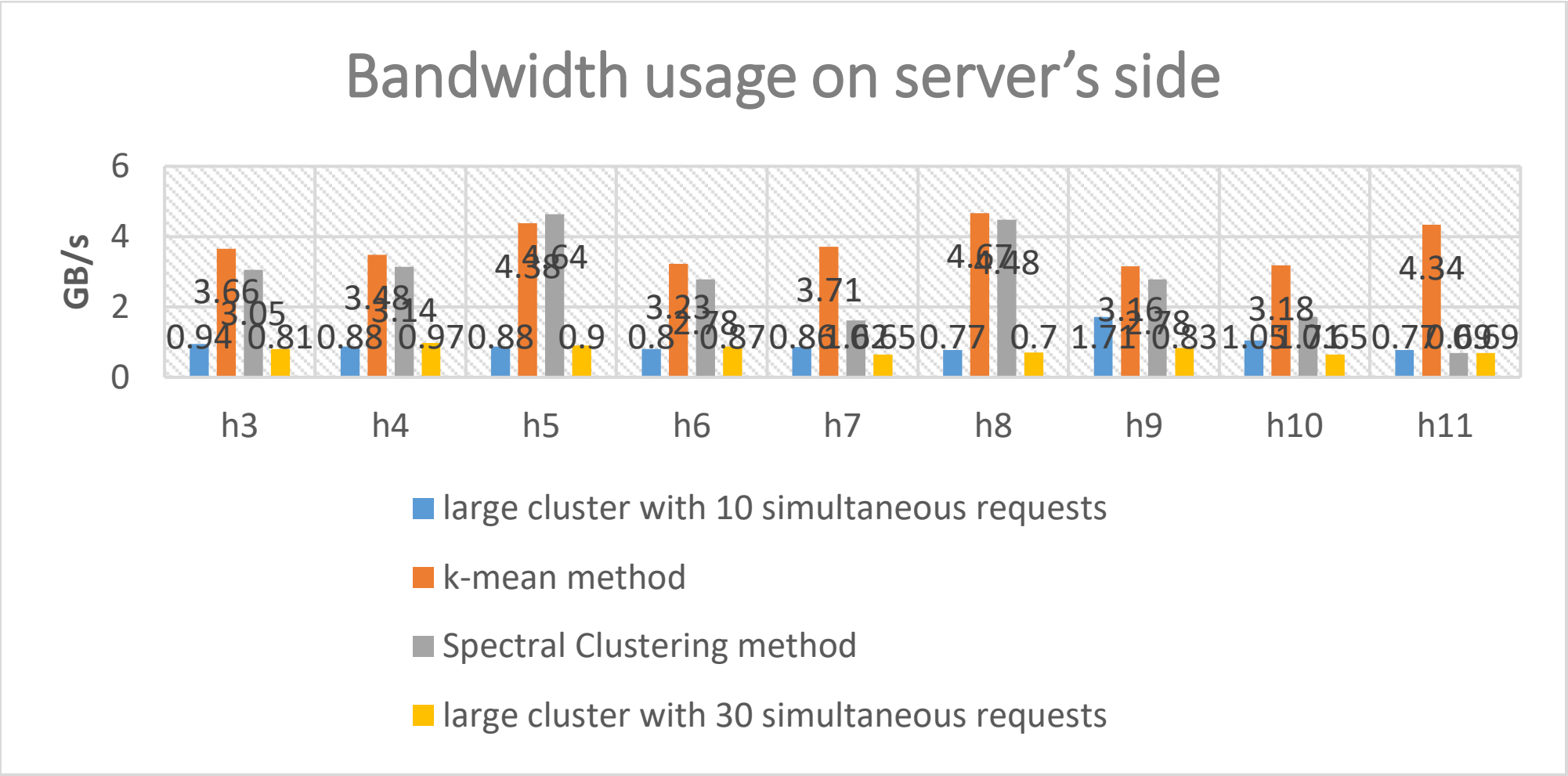

Figure 8. Bandwidth usage on server's side

Figures 9 and 10 shows transmitted data volume and occupied bandwidth in user's side in 10 seconds.

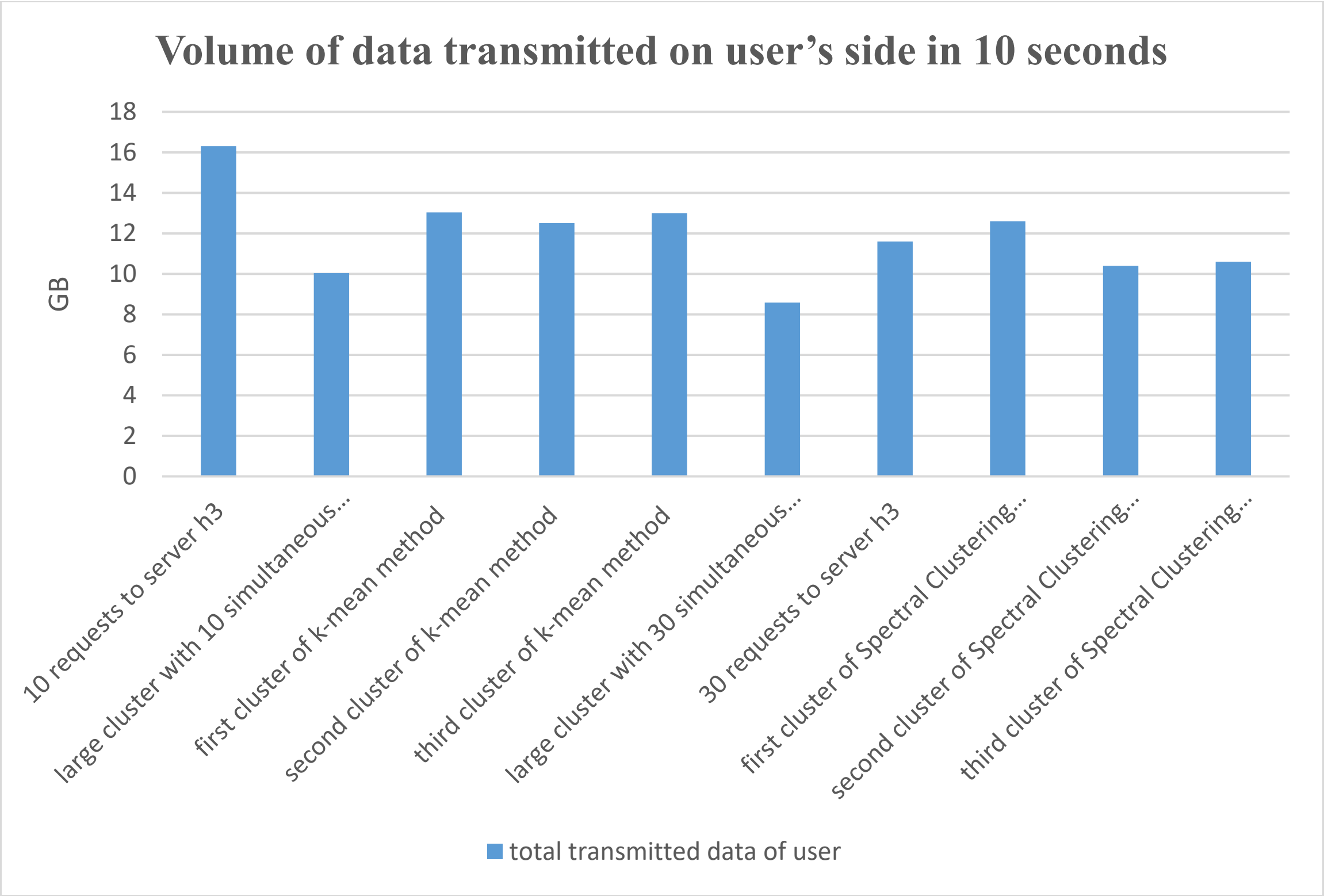

Figure 9. Volume of data transmitted on user’s side in 10 seconds

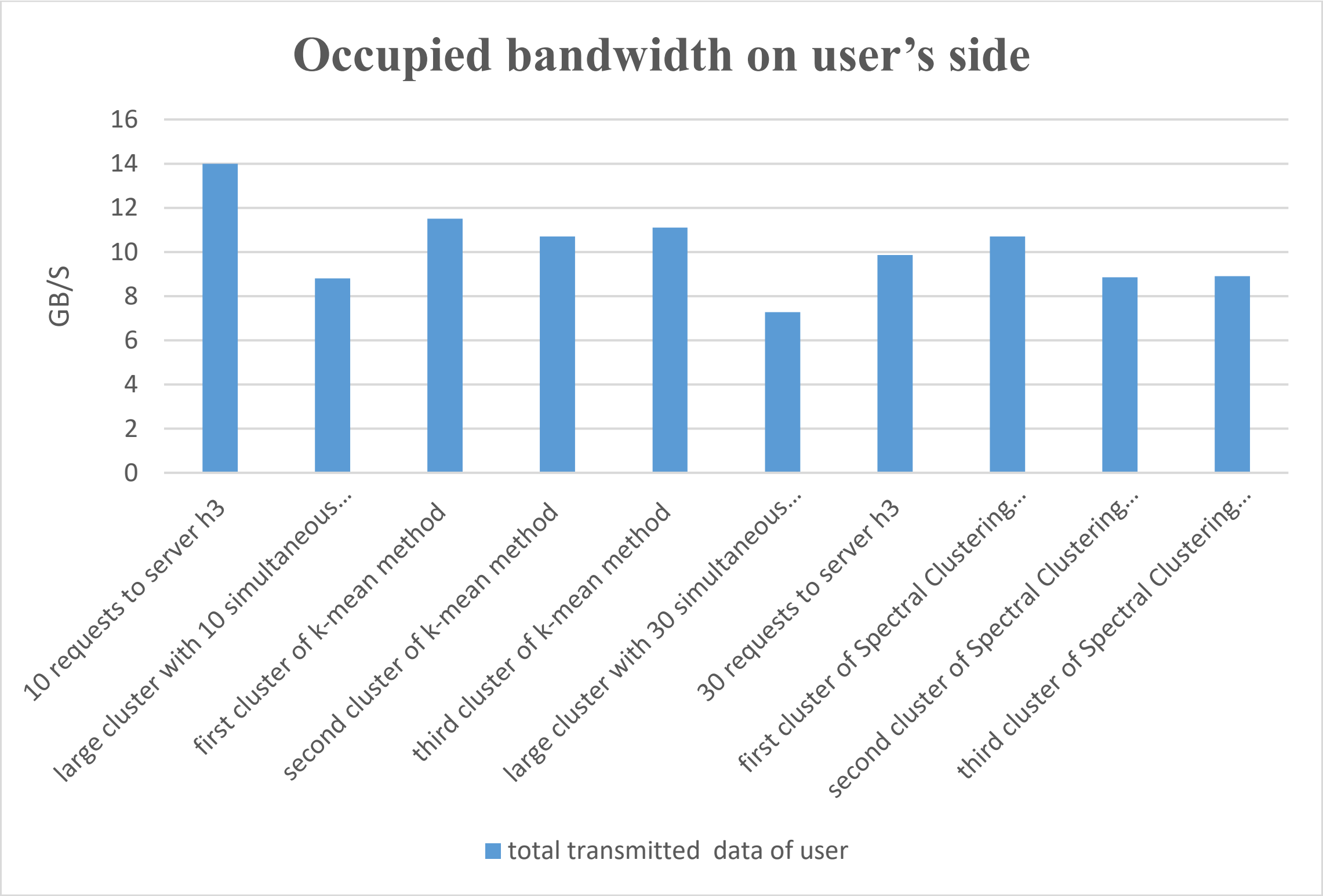

Figure 10. Occupied bandwidth on user’s side

As can be seen in the figures 9 and 10 diagrams, in the proposed method, volume of data transmitted in each server and user's side has increased several times compared to the large cluster and more bandwidth is used. More bandwidth usage without allocating any new resource indicates increase in throughput and significant improvement. As can be seen on user's side, each cluster experiences a significant improvement compared to the large cluster and in the proposed method, more data is transmitted in 10 seconds. Indeed, when number of requests is low, each cluster performs better separately considering load balancing complexity of each server. But as number of requests increases, a large drop is experienced and the proposed method performs better considering that 30 requests are divided into three groups including 10 requests.

In comparison of clusters, as can be seen in figure 3, first cluster has the highest transmitted data and maximum bandwidth usage because it is the closest cluster and middle cluster or the third cluster has lower quality compared to the first cluster and farthest cluster or second cluster has minimum transmitted data. Considering the above order, quality of service can be specified considering clustering for each cluster.

## 5. Conclusion

In this study, a novel approach is proposed for improving storage resource allocation in network using load distribution clusters. Since in the proposed method, k mean method is used for clustering, network load balance is preserved by preserving load balancing of clusters and by allocating high quality clusters in terms of average number of hubs and average delay between server and user, quality of service and network efficiency can be increased for data with higher recall. On the other hand, increasing number of clusters is directly related to storage capacity of the network and by increasing number of cluster, network storage can be increased. The proposed method can be used as a preprocessing in networks to store high-volume data. In addition, considering priority of clusters, users can be classified and provide each class with desired quality level.

## 6. Future works

Since SDN context is new, each challenge of this context can be used as a research topic. In this study, load balancing in the network an optimal resource allocation are considered. In the proposed method, resources are clustered using k-means algorithm. It is suggested to use other clustering method for improving clustering in future works. Furthermore, other data-mining methods can be used instead of clustering.